\documentstyle[12pt]{article}
\begin{document}
\renewcommand{\thesection}{\arabic{section}.}
\newcommand{\tq}{top quark }
\newcommand{\nn}{\nonumber}
\newcommand{\nl}{\newline}
\newcommand{\gp}{g^{\prime} }
\newcommand{\gpp}{g^{\prime\prime} }
\newcommand{\be}{\begin{equation}}
\newcommand{\ee}{\end{equation}}
\newcommand{\elw}{electroweak }
\newcommand{\bea}{\begin{eqnarray}}
\newcommand{\eea}{\end{eqnarray}}
\newcommand{\lb}{\label}
\newcommand{\lp}{\Lambda_{Planck}}
\newcommand{\lc}{\Lambda_c}
\newcommand{\pr}{\prime}
\newcommand{\al}{\alpha}
\newcommand{\ga}{\gamma}
\newcommand{\half}{\frac{1}{2}}
\newcommand{\sqrthalf}{\frac{1}{\sqrt{2}}}
\newcommand{\gt}{g_{top}}
\newcommand{\de}{\delta}
\newcommand{\De}{\Delta}
\newcommand{\ti}{\times}
\newcommand{\ra}{\rightarrow}
\newcommand{\lra}{\longrightarrow}
\newcommand{\mi}{\mbox{i}}
\newcommand{\mtr}{\mbox{Tr}\,}
\newcommand{\mt}{m_{top}}
\newcommand{\smgauge}{SU(3)_c \ti SU(2)_L \ti U(1)}
\newcommand{\xxg}{SU(2)_L \ti U(1) \ti G}
\newcommand{\vu}{v_u}
\newcommand{\vd}{v_d}
\renewcommand{\thefootnote}{\fnsymbol{footnote}}
\begin{titlepage}
\begin{quote}\raggedleft  {\bf September 94} \end{quote}
\vspace{4cm}\begin{center}
{\bf Electroweak Condensates at Standard Top Mass Values}
\footnote{
Invited Talk given at the XVII Kazimierz Meeting on Elementary
Particle Physics "Facing the Desert or New Physics",
Kazimierz, Poland, May 23-27, 1994 and at the
IX International Workshop "High Eenergy Physics and Quantum Field Theory",
Zvenigorod, Moscow Region, Russia, September 16-22, 1994. To appear in the
latter proceedings.}
\normalsize
\end{center}
\vspace{1cm}
\begin{center} {R. B\"onisch}
\\
 DESY-IfH, Platanenallee 6, D-15738 Zeuthen, Germany
\end{center}
\vspace{4cm}
\begin{abstract}
In the two Higgs-doublet SM, perturbativity
breaks down below $\lp$: For $\mt = 170 - 180$ GeV, $\gt$ reaches
the pole at scales below 1000 TeV if the ratio of {\sc vev}'s $\vu/\vd$
is smaller than 1.
We discuss top condensate scenarios
assuming 150 GeV $\leq \mt \leq$ 200 GeV:
The phenomenological work
of the last years shows a version with
two composite Higgs doublets as a viable option. In the theoretical
frame of gauge extensions, a second doublet does not come with additional
parameters, but is dictated by the symmetry.
\end {abstract}
\end{titlepage}
\renewcommand{\thefootnote}{\fnsymbol{footnote}}
Top condensate models have been motivated a few years ago when
the top mass $\mt = \sqrthalf \gt v$ became known to be as heavy as
60 or 70 GeV \cite{Nambu}. With such masses,
top is the only elementary fermion
with $\mt/M_W = {\cal O}(1)$ and
this - rather than the fact that the top
is heavier than all remaining eleven fermions together -
is what makes the top quark special: the dynamics of $\mt$ and $M_W$
must be related.
The minimal, effective model replaces the standard Higgs-sector
by an NJL interaction involving $t$ and $b_L$,
\be G \bar{\Psi}_L^i t_{iR}
\bar{t}_R^j \Psi_{jL}, \hspace{.7cm} \Psi=
\left( \begin{array}{c} t \\ b  \end{array}\right),\hspace{.7cm}i,j=1,...N,
\lb{bhl}\ee
which gives rise to a composite $\sigma$-model if $G>G_c=8\pi^2/N$.
The scalar states are a massive $\bar{t}t$ and massless $\bar{t}\ga_5t$,
$\bar{b}_Lt_R$, $\bar{t}_Lb_R$. The top has no bare mass term and achieves
a dynamical one via a self-consistent Schwinger-Dyson summation.

Because we now have growing evidence for $\mt \simeq 174$  GeV from CDF
\cite{CDF}, it is useful
to take a fresh look on this well motivated kind of model
under the assumption
\be 150 \,\, {\mbox GeV} < \mt < 200 \,\,{\mbox GeV}.\lb{topass} \ee
It will be useful to follow the discussion of the model
like it proceeded in the major part of the
literature within the last 5 years, because
the model variations left first statements mainly intact
and numerical changes of
predictions can be derived directly
from the extension or variation under consideration.
The result will be that all such numerical changes are small
for viable parameter ranges, i.e.
the prediction of the minimal model is (maybe surprisingly) stable in
extensions.
What has often been regarded as a failure to construct models
with the correct $\mt$ prediction now turns out to be a
confirmation of the two composite Higgs-doublet case:
Models with one Higgs-doublet predict $\mt > 200 GeV$,
while a two Higgs-doublet
model predicts $\mt$ in the range (\ref{topass}) with a condensation scale
possibly as low as a few TeV.

The pure NJL prediction is given by the relation $m_H=2 \mt$, but
$\smgauge$ gauge interactions must be included.
They have been accounted for in solving
the SM RG-equation for $\gt$ with the boundary
condition $\gt (\mu \ra \Lambda_c) \ra \infty$, where $\Lambda_c$ is the
condensation scale at which the composite scalar field
$ \bar{t}t = Z(\mu)\phi_0$
vanishes with $Z=g_{top}^0/\gt$ \cite{bhl}.
This is practically the same as considering the running of $\gt$ in
the minimal SM itself and looking for the Landau pole.
The use of 1-loop RG-equations in this procedure is believed to be a
meaningful approximation, because the breakdown of perturbativity takes place
not too far below $\Lambda_{Planck}$ \cite{Kim+Cvetic} and numerical
calculations support the results \cite{Carena+Wagner}.
It is well-known that in the minimal model, $\mt$ comes out too large:
values are above the window allowed by the $\rho$-parameter even for very
highly tuned
$\Lambda_c > 10^{16}$ GeV. This is saying that the pole is not
reached before $\lp$ if $\mt < 200$ GeV.
Thus we are confronted with an extreme (and in fact a worse case) of
the unnatural Standard desert situation, as all quadratic
divergencies have only been absorbed into the top mass counterterm.

The situation drastically changes, if the scalar sector possesses two
doublets: the {\sc vev}'s of up- and down-type Yukawa couplings
add up in the $W$ self-energy and therefore have to
satisfy $v^2 = \vu^2 + \vd^2$. This simply causes a lower $v$ to enter
the fermion mass $\sqrthalf g_f v$ so that $g_f$ has to be fitted
to a higher value. The influence can be read off best from table 1 by Kim
and Cvetic \cite{Kim+Cvetic}.
For values (\ref{topass}) the breakdown is then reached at
comfortable scales 1-1000 TeV, if $\vu/\vd < 1$.
Such ratios enhance the running of $\gt$. The same was
found in the
supersymmetric version of the TMSM \cite{susy}, where the second doublet
and quadratic divergencies appear in a different context.

The second serious objection against the minimal TMSM, put forward
mainly by A. Hasenfratz et al., concerned
predictability \cite{universality}:
The pure effective NJL terms induce all sorts of higher
dimensional counterterms.
Among them are derivative couplings, which correspond to
momentum dependent loop diagrams like Fig.1, \cite{Suzuki}. These terms
represent the propagation of the bound state (the internal loop) and
inclusion of the complete series of derivative couplings, each with a
free coefficient, leads back to the SM universality class.
Having this critics only qualitatively in mind, gauge interactions
were used to yield the NJL operator on the Fermi scale as
a well-defined part of a complete gauge boson exchange:
The new parameters
are a gauge coupling $g_{new}$, a degree of freedom $N$ to expand in and
a new scale $\lc \ll \lp$,
all of
which have definite physical meaning, such that we now gain information
on physics beyond the SM out of the composite Yukawa sector
\cite{Bonisch}.
Higher dimensional terms are calculable and small for viable ranges of $\lc$
(second ref. \cite{gauge}).

The gauge extension causes deviations from the pure NJL prediction and we
want to classify the new effects in a model-independent way as far as
possible.
The new boson(s)
shall be strongly coupled and heavy. It is useful to expand
the new interaction in powers of $p^2/M^2$, where
$M \sim \lc$ is the new mass and $p$ a typical momentum and write
the complete form as:
\be {\cal L} = J_L^2 + J_R^2 + J_LJ_R + J_RJ_L +{\cal O}(p^2/M^2),
\lb{full}\ee
\be  J_LJ_R + J_RJ_L = \sum_{i,j=1}^N \frac{G_{LR}}{M^2} \left(
\bar{\psi}_L^i \psi_{iR} \right) \left(
\bar{\psi}_R^j  \psi_{jL} \right)   +h.c.\nn \ee
\bea
J_L^2&=& \sum_{i,j=1}^N \frac{G_{LL}}{M^2} \left(
\bar{\psi}_L^i \ga^\mu \psi_{iL} \right) \left(
\bar{\psi}_L^j \ga_\mu \psi_{jL} \right),    \nn \\
J_R^2&=&     \sum_{i,j=1}^N \frac{G_{RR}}{M^2} \left(
\bar{\psi}_R^i \ga^\mu \psi_{iR} \right) \left(
\bar{\psi}_R^j \ga_\mu \psi_{jR} \right),
\eea
Here we can see that
the model is not in one universality class with the SM:
$J_{L(R)}^2$ and ${\cal O}(p^2/M^2)$ are new terms which
cannot be decoupled from weak physics,
because tuning the new physics to scales
$\Lambda_c \gg M_W$ must keep $M_W$ itself fixed.
This can only be done by simultaneously tuning $G_{LR} \ra G_c$ and this
takes place in every channel of eq. (\ref{full}) as $G_{LR} \sim G_{LL(RR)}$.
Additional predictions
from these additional operators will
remain at low energies and there is no way to reach the minimal
SM as a limit. Let us now recall the known results
from those non-minimal predictions/effects,
beginning with $\rho$:

The Fierz-eigenstates $J_L^2$ and $J_R^2$
contribute to the bubble summation which corrects $\rho$, Fig. 2.
They have been calculated to ${\cal O}(1/N)$ for $p^2=0$
and found to be negligible \cite{Bonisch+Kneur}.

The next important part that contributes to $\rho$ is the
vector resonance spectrum. It is of course completely model-dependent
and constitutes the yet unknown part of new effects in $\rho$.
Vector resonances will show non-decoupling as discussed above.
On the other hand, one
expects these resonances to be uncritical, i.e. although the tuning
enters in the corresponding channel, divergencies remain as no
gap-equation automatically removes them. Therefore they shall
be of ${\cal O}(\Lambda_c)$ and the decoupling takes place for this
mass to suppress the influence on
observables at low energies (like especially $\rho(0)$).

Using a dynamical function $\Sigma_{top}(p^2)$ in the $W$-self-energy,
A. Blumhofer and M. Lindner argued that a cancelation of the SM $\mt^2$
term in $\rho$ can take place
if $\Sigma_{top}$ has some kind of resonant enhancement at $\sim 5 \mt$
\cite{Lindner+Blumhofer}.

A further effect comes from ${\cal O}(p^2/M^2)$ terms in eq.(\ref{full}).
They represent the $p$-dependent
short distance parts of the propagator of new bosons and thus
easily cause additional
resonances (radial excitations of composite states).
M. Lindner and D. L\"ust have considered additional scalars and vectors in the
SM-RG-eq. \cite{Lindner+Luest}.
The running of $\gt$ is altered by a change of slope at the scale
of the resonance. Additional scalars (vectors)
result in a decreasing (increasing) $\mt$ and
one additional scalar at 100 GeV
in the otherwise minimal SM
for example allows $\mt < 200 $ GeV at $\Lambda_c \sim 10^{11}$ GeV.
Altogether, the ${\cal O}(p^2/M^2)$ terms do not drastically change
predictions if not a number of scalars form around
$M_W$. This requires $M$ to be of ${\cal O}(M_W)$, a too low value.

In all these examples above non-standard effects are sizeable only
if $M (\sim \lc) \stackrel{<}{\sim} $
1 TeV. The analysis of non-standard $p_T$ in
$\bar{t}t$ production has not been shown to give a different
result \cite{Hill}. We should now look for lower bounds on $M$.

A $Z^\pr$ mass is of course limited already in precision measurements.
Perturbative bosons are usually bounded to be heavier than a few hundred
GeV, depending on the details of the model which give the physical couplings
of the various precisely measured channel
\cite{zpr}. For a strongly coupled $Z^\pr$,
the limit depends on the effective fermionic coupling
$g^2_{new}=G$ of $Z^\pr$, which
has to exceed the critical coupling $8\pi^2/N$. The limit thus depends on
$N$.
A strong coupling at LEP100, assuming $N=3$, gives a typical limit
$M_{Z^\pr} > $ 3 TeV \cite{Bonisch+Leike}, Fig. 3.
If no new bosons are found at LEP200
or a 500 GeV $e^+e^-$ collider, the limit increases to 15 TeV and 50 TeV
respectively, Fig. 4.
This analysis uses a static coupling and correct running of $g_{new}$
in the critical channels can cause drastic changes, non-critical
channels will however remain to be strong at low energies.

Collecting the above points, we see that i)
augmenting the interaction eq.(\ref{bhl}) to
eq.(\ref{full}) does not significantly
effect the result of the pure NJL gap-equation
or standard RG-running unless $\Lambda_c \stackrel{<}{\sim}$ 1 TeV
and ii) such scales seem rather unlikely from experiment already.
Turning this around, the NJL-RG prediction is rather stable
to hold in a gauge extension at an expected
\be \Lambda_c \stackrel{>}{\sim} \mbox{1 TeV}.
\lb{1tev}\ee
This disfavors the one Higgs-doublet
model under the assumption (\ref{topass}), while the
two doublet model can live well with (\ref{topass}) and (\ref{1tev}):
A lack of non-standard negative contributions to low energy
observables like $\rho$ in this scenarios
preserves indirect
measurements of $\mt$ to contradict with the NJL-RG
result (table 1).

Let us make a remark on a previously proposed type of model.
The empirical scaling formula for masses $m_i$ across flavor components
$i=1,2,3$,
${m_2}/{m_1}=3 \left( {m_3}/{m_2}
\right)^{\frac{3}{2}|Q|}$, \cite{Sirlin},
tells us that $Q$ can be used for the mass splitting. The most
straightforward
way to involve $Q$ in the top condensate model is to couple the new boson
to hypercharge currents by mixing in $SU(2)_R$ or any other hidden
factor of the SM \cite{Schildknecht,Bonisch}.
The scalar couplings $G_{LR}$, relevant in the SD-eq.,
are proportional to
\be Y_{L} \cdot Q= \left\{
\begin{array}{ll}\frac{1}{6}\cdot\frac{2}{3} &\mbox{for $I_{W_3}=1/2$
quarks, }\\
 \frac{1}{6}\cdot(-\frac{1}{3}) &\mbox{for $I_{W_3}=-1/2$
quarks.}
\end{array} \right . \lb{quarkhier}\ee
The mass matrix is of rank 1 in
generation space.
According to eq. (\ref{quarkhier}) only $I=+1/2$ quarks condense, while
the interaction is repulsive for $I=-1/2$ quarks, i.e. interaction
eq. (\ref{bhl}) is recovered and only the top
is massive.
There is an analogous term
for the leptons,
$
\bar{\Psi}_L \tau_R \bar{\tau}_R \Psi_L, \,\,
\Psi =(\nu_{\tau},\tau)^T $,
which among leptons produces only a mass for the $\tau$. This channel
is stronger than the top-bottom system,
$ Y_{L} \cdot Q= -
\frac{1}{2}\cdot (-1)$, for $I_{W_3}=-1/2$
leptons,
possibly leading to the required situation $v_{u} / v_{d} <1$.
Additionally, there is one more attractive channel
involving $b_R$ and $(\nu_\tau, \tau)_L$, intermediate in strength and
giving rise to a system of coupled gap-equations, which has not been
considered yet.
A generalization of the representation was introduced in \cite{Bonisch2}.

We note that there
is further work on condensate type models, both without the derivation
of new interactions \cite{Gribov}  or
scenarios like the inclusion of a 4th family
\cite{4th} and other variations, which do not fall into special
classes of models (see \cite{gauge} and further refs. therein).
\vspace{.5cm}\nl
I wish to thank
 S. T. Pokorski and Z. Ajduk for the kind invitation to and
 hospitality at the Kazimierz meeting and
the Organizing Committee
in Zvenigorod very much for the kind arrangements that made the visit a
real pleasure.

%\pagebreak[4]

\end{document}